\documentclass[aps,prb,twocolumn,groupedaddress]{revtex4}
\usepackage[dvips]{graphics}
\usepackage{vmargin}
\setpapersize{USletter}
\setmarginsrb{0.8in}{.8in}{0.8in}{0.8in}{0in}{0in}{0in}{0in}
\begin{document}

\title{Localization-delocalization transition in 2D quantum percolation model}
\author{Md Fhokrul Islam}
\author{Hisao Nakanishi}
\affiliation{Department of Physics, Purdue University, West
Lafayette, IN47907}
\date{\today}

\begin{abstract}
We study the hopping transport of a quantum particle through randomly diluted
percolation clusters in two dimensions realized both on the square and triangular
lattices. We investigate the nature of localization of the particle by calculating
the transmission coefficient as a function of energy ($-2 < E < 2$ in units of the
hopping integral in the tight-binding Hamiltonian) and disorder, q (probability
that a given site of the lattice is not available to the particle).
Our study based on finite size scaling suggests the existence of delocalized states
that depends on energy and amount of disorder present in the system. For energies away
from the band center ($E = 0$), delocalized states appear only at low disorder
($q < 15\%$). The transmission near the band center is generally very small for
any amount of disorder and therefore makes it difficult to locate the transition
to delocalized states if any, but our study does indicate a behavior that is weaker
than power-law localization. Apart from
this localization-delocalization transition, we also find the existence
of two different kinds of localization regimes depending on energy and amount of
disorder. For a given energy, states are exponentially localized for sufficiently
high disorder. As the disorder decreases, states first show power-law localization
before showing a delocalized behavior.

\end{abstract}
\maketitle

\section{Introduction}
Unlike a classical particle, the transport of a quantum particle
through a system is greatly influenced by the interference and
tunneling effects. While the availability of a spanning path is the
sole criteria for the transmittance of a classical particle through
a system, a quantum particle may exhibit zero or very low
transmittance even for a completely ordered system depending on
details such as the boundary condition or the energy of the particle
\cite{cuansing:04}. The quantum percolation system that we have
investigated here includes the interference effect but does not
include the tunneling effect. We, thus, expect a higher connectivity
in underlying geometry to be required for non-zero transmission
compared to its classical counterpart.

A major motivation for studying such a system is the question of whether a
localized-to-delocalized (or perhaps, metal-to-insulator) transition exists
in a two-dimensional (2D) system. The Anderson model and the quantum percolation
model are two of the more common theoretical models that are used to study the
transport properties of disordered systems. While the literature on both models
agree on the existence of such a transition in three dimensions
\cite{souk:87,kosl:91,berk:96}, the same question
for quantum percolation in two dimensions appears to have remained a subject of
controversy for over two decades. Based on the one-parameter scaling theory of
Abrahams et al \cite{abrahams:79}, it was widely believed that there can be
no metal-to-insulator transition in 2D universally in the absence of a magnetic field
or interactions for any amount of disorder. Moreover, the scaling theory predicts
that all states are exponentially localized in thermodynamic limit for any amount
of disorder and therefore that no transition exists except at zero dilution. (However,
see Goldenfeld and Haydock\cite{goldenfeld:06} which asserts the existence of
a transition between two different kinds of localized regimes at a finite disorder
in addition to a localized-delocalized transition at an infinitesimal dilution,
even for the Anderson model in two dimensions.)

However, whether the scaling theory also applies to quantum percolation
has been debated in recent years. Even restricting attention to quantum
percolation which lacks many effects that are expected to play important roles in
metal-insulator transitions, there is a long-standing controversy as to the presence
or absence of an extended state and of a phase transition between the prevalent
localized state and a more elusive extended state in two-dimensions. On one hand,
some studies such as those made using the dlog Pad\'{e} approximation method
\cite{daboul:00}, real space renormalization method \cite{odagaki84}, and the inverse
participation ratio \cite{srivastava84} found a transition from exponentially
localized states to non-exponentially localized states for a range of site
concentrations between $0.73 \leq p_q \leq 0.87$ on the square lattice. So did
a study of energy level statistics \cite{letz:99}, one of the spread of a wave
packet initially localized at a site \cite{nazareno:02}, and one of a transfer
matrix \cite{eilmes:01}, where the nature of the delocalized state remained
not fully understood. On the other hand, studies such as the scaling work based
on numerical calculation of the conductance \cite{haldas02}, the investigation of
vibration-diffusion analogy \cite{bunde98}, finite-size scaling analysis and
transfer matrix methods \cite{soukoulis91}, and vector recursion technique
\cite{mookerjee95} found no evidence of a transition.  A study by Inui {\it et al.}
\cite{inui94} found all states to be localized except for those with particle energies
at the middle of the band and when the underlying lattice is bipartite, such as
a square lattice. More recently, Cuansing and Nakanishi \cite{cuansing:06}
used an approach first suggested by Daboul {\it at al.} \cite{daboul:00} to calculate
conductance directly for clusters of up to several hundred sites and, extrapolating
those results by finite-size scaling, suggested that delocalized states exist
and thus a transition would have to exist as well. The current paper extends
the relevant portion of the latter work by studying much larger clusters, which
has allowed considerably more detailed analyses.

In the mean time, experiments performed in early 1980's on different 2D systems
\cite{exp:1,exp:2,exp:3} confirmed the scaling theory predictions. However, a
number of experiments on dilute low disordered Si MOSFET and GaAs/AlGaAs
heterostructures that appeared more recently seem to suggest that a metallic
state may be possible in two dimensions \cite{Krav:94,Krav:95,Sara:99,Sara:01}.
For reviews of these experiments, see Abrahams {\it et al.} \cite{abrahams:01}
and references therein.

In this work, we do not address the issues of these experiments, but rather
concentrate on the formally much simpler quantum percolation model which has
neither magnetic field nor interactions but contains binary disorder with
infinite barriers at randomly diluted sites. Previously we have investigated
the same problem using a dynamical approach where we have studied the properties
of a disordered system by tracking how a quantum particle, described by a wave
packet, propagates through the system \cite{islam1:07}. In this paper we
adopt a stationary state approach where we calculate the transmission
characteristics by solving time independent Schr\"{o}dinger equation for
tight binding Hamiltonian. In Section II we describe the model and the numerical
approach used in this work. In Section III we discuss the numerical results
and in Section IV we present the summary and conclusion of our study.

\section{Quantum Percolation Model and Numerical approach}

We study quantum percolation that is described by the Hamiltonian
\begin{equation}
H = \sum_{<ij>} V_{ij}|i\rangle \langle j| + h.c
\label{eq1}
\end{equation}

where $|i\rangle$ and $|j\rangle$ are tight binding basis functions at sites $i$ and $j$,
respectively, and $V_{ij}$ is the hopping matrix element which is equal to zero
if $i$ and $j$ are not nearest neighbors. We have realized this model on both square
and triangular lattices that can have at most 4 and 6 nearest neighbors, respectively.
If the system is completely ordered, $V_{ij} \equiv V_{0}$ (uniform) and $V_{0}$ sets
the overall energy scale, where we use $V_{0} = 1$ as the nominal standard value.
On the other hand, since in this work we are interested in transport through a
disordered system, we will introduce random dilution by removing a fraction of sites
from the lattice and set $V_{ij} = 0$ for the bonds between the diluted sites and their
neighbors. $V_{ij} = 0$ for nearest neighbors $i$ and $j$ means that an infinite barrier
exists between the pair of sites.

\begin{figure}[htbp]
{\resizebox{3.5in}{2.7in}{\includegraphics{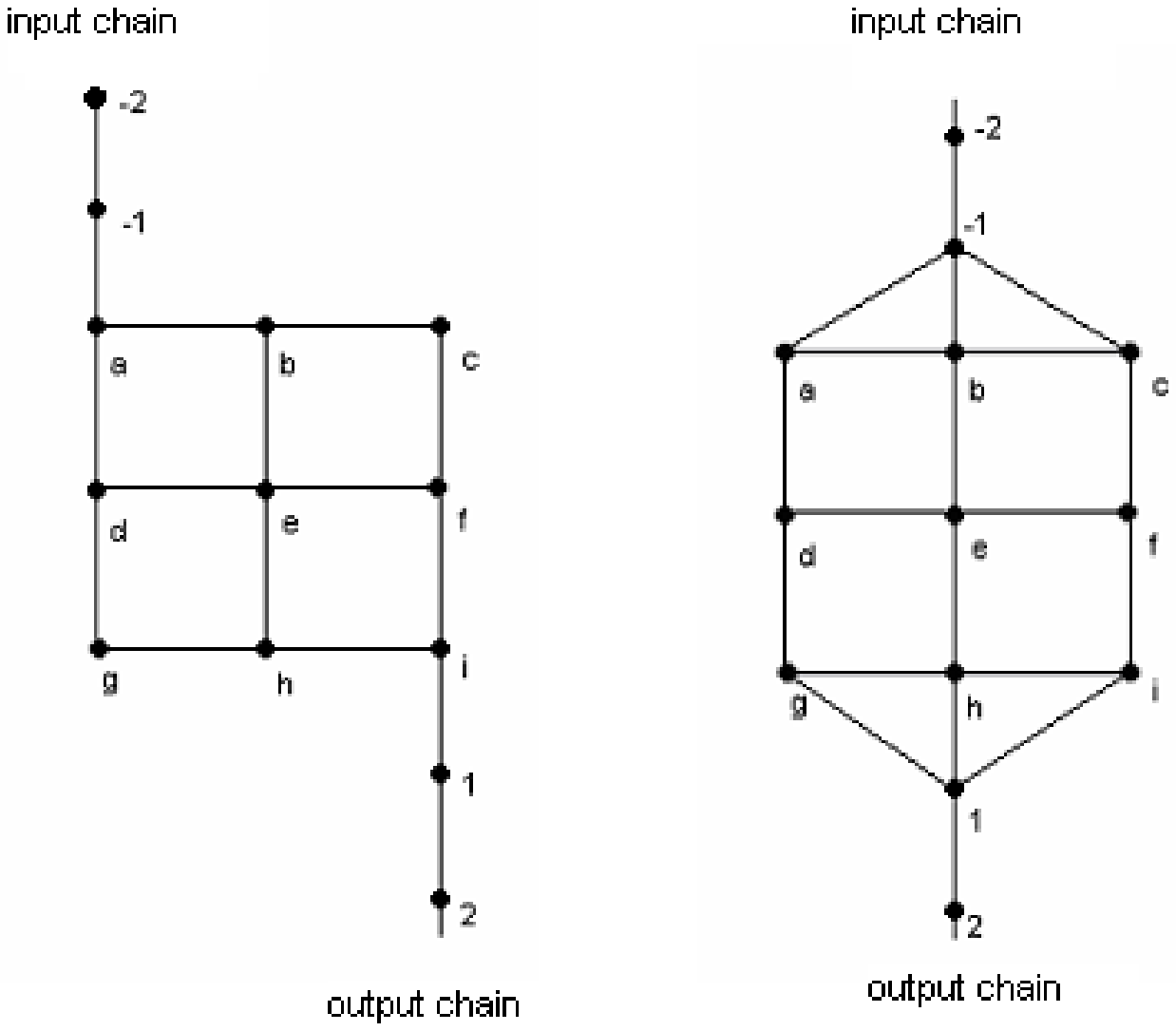}}}\\
(a) \hspace{1.5in} (b)
\caption{$3 \times 3$ Square Lattice: (a) point-to-point connection and (b) busbar type connection.
          The letters label the lattice points of the cluster part of the Hamiltonian, while
          numbers label those of the leads. The same sequence of labeling is used for all sizes
          of the clusters in this work.}
\label{sq_lat}
\end{figure}

\begin{figure}[htbp]
{\resizebox{3.5in}{2.7in}{\includegraphics{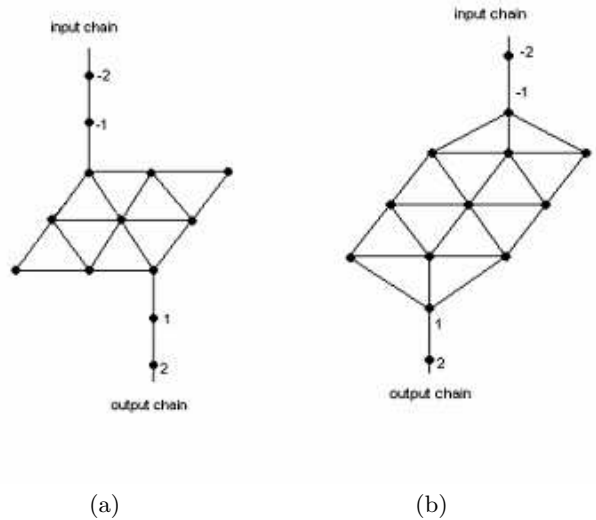}}}
(a)\hspace{1.5in} (b) \caption{$3 \times 3$ Triangular Lattice: (a)
point-to-point connection and (b) busbar type connection}
\label{tri_lat}
\end{figure}

To study the transmission of a quantum particle we connect two semi-infinite 1D leads,
one as the input and the other as the output lead, to the 2D cluster. Although the system
can be studied using different types of connection of the leads, in this study we
only use a point-to-point type contact where the input lead is connected to only one
lattice site on the input side edge of the cluster and the output lead is connected
also to only one lattice site on the opposite edge of the cluster. Another possible
connection type is the busbar type contact, where all the lattice points on
the input side edge of the cluster are connected to the input lead, while all the
lattice points on the output side of the cluster are connected to the output lead.
Figs.~\ref{sq_lat} and ~\ref{tri_lat} illustrate the connection of the leads for
the square and the triangular lattices respectively.

The wave function of the entire cluster-lead system can be calculated by solving the time
independent Schr\"{o}dinger equation:

\begin{equation}
\begin{array}{l}
H\psi = E\psi  \\
\mbox{where}, \psi = \left( \begin{array}{c} \psi_{in} \\ \psi_{cluster} \\ \psi_{out} \end{array}
               \right)
\end{array}
\label{eq2}
\end{equation}

and $\psi_{in} = \left\{\psi_{-(n+1)}\right\}$ and $\psi_{out} = \left\{\psi_{+(n+1)}\right\}$,
$n = 0,1,2 \ldots$, are the input and output chain part of the wave function respectively.\\

Since the leads are of infinite length, the matrix form of the Schr\"{o}dinger equation
(Eq.~\ref{eq2}) becomes an infinite size problem. To reduce it to a numerically finite
problem we use an ansatz proposed by Daboul {\it et al.} \cite{daboul:00} which assumes
that the input and output part of the wave function are of the form of plane waves:

\begin{equation}
\begin{array}{l}
\psi_{in} {\rightarrow} \psi_{-(n+1)} = e^{-inq} + re^{inq} \\
\psi_{out} {\rightarrow} \psi_{+(n+1)} = te^{inq}
\end{array}
\label{eq3}
\end{equation}

where {\it r} is the amplitude of reflected wave, {\it t} is the amplitude of the
transmitted wave. This ansatz is consistent with the Schr\"{o}dinger equation only
for the wave vector $q$ that is related to the energy {\it E} by

\begin{equation}
E = e^{-iq} + e^{iq}
\label{eq4}
\end{equation}

Using this ansatz along with the energy restriction Eq.~(\ref{eq4}), the matrix
equation for a $3 \times 3$ cluster connected to semi-infinite chains
(Fig.~\ref{sq_lat}a) reduces to (for details see reference \cite{daboul:00})

\begin{widetext}
\begin{equation}
\left( \begin{array}{crrrrrrrrrc}
         -E + e^{iq} & c  &  0 &  0 &  0 &  0 &  0 &  0 &  0 &  0 &    0 \\
               c     & -E &  1 &  0 &  1 &  0 &  0 &  0 &  0 &  0 &    0 \\
               0     & 1  & -E &  1 &  1 &  1 &  0 &  0 &  0 &  0 &    0 \\
               0     & 0  &  1 & -E &  0 &  0 &  1 &  0 &  0 &  0 &    0 \\
               0     & 1  &  0 &  0 & -E &  1 &  0 &  1 &  0 &  0 &    0 \\
               0     & 0  &  1 &  0 &  1 & -E &  1 &  0 &  1 &  0 &    0 \\
               0     & 0  &  0 &  1 &  0 &  1 & -E &  0 &  0 &  1 &    0 \\
               0     & 0  &  0 &  0 &  1 &  0 &  0 & -E &  1 &  0 &    0 \\
               0     & 0  &  0 &  0 &  0 &  1 &  0 &  1 & -E &  1 &    0 \\
               0     & 0  &  0 &  0 &  0 &  0 &  1 &  0 &  1 & -E &    c \\
               0     & 0  &  0 &  0 &  0 &  0 &  0 &  0 &  0 &  c & -E + e^{iq}
          \end{array} \right)
          \left( \begin{array}{c}
                    1 + r \\ \psi_{1} \\ \psi_{2} \\ \psi_{3} \\ \psi_{4} \\ \psi_{5} \\ \psi_{6}\\
                    \psi_{7} \\ \psi_{8}\\ \psi_{9} \\ t
                    \end{array} \right)
         = \left( \begin{array}{c}
                  e^{iq} - e^{-iq} \\ 0 \\ 0 \\ 0 \\ 0 \\ 0 \\ 0 \\ 0 \\ 0 \\ 0 \\ 0
                  \end{array} \right)
\label{eq5}
\end{equation}
\end{widetext}

Here $c$ is the coupling of the leads with the cluster and its value is set to 1 for
all the calculations done in this work. The effect of $c \neq 1$ on transmission
is discussed in one of our previous works \cite{islam2:07}. The busbar configuration
of Fig.~\ref{sq_lat} (b) has a similar expression.

The Eq.~(\ref{eq5}) is the exact expression for a 2D system connected to semi-infinite
chains with continuous eigenvalues ranging between -2 and +2. The spectrum is
continuous because it is still effectively infinite and it is non-degenerate except
for the reversal of left and right.

The main advantage of using this ansatz is that it not only allows us to calculate
the wave function but also helps us to study the transmission characteristics of
the corresponding state directly. The transmission and the reflection coefficients
are obtained by taking the absolute square of {\it t} and {\it r} respectively, ie
$T = |t|^{2}$ and $R = |r|^{2}$. A disadvantage on the other hand is that
Eq.~(\ref{eq4}) that relates the wave vector of the incident particle with the energy,
restricts the energy of the particle to between -2 and +2. This restricts our ability
to study the system in the whole possible energy range since for our system the energy
could in principle range between -4 and +4. This is due, of course, to the effectively
one-dimensional nature of the system forced by attaching 1D semi-infinite leads and
by looking at plane waves spreading over the entire leads.

\section{Numerical results}

The model that we are using to calculate transmission has two adjustable parameters, namely
the energy of the particle, E and the amount of disorder present in the system, q
(probability that a given site is not available to the hopping particle).
To study the presence or absence of a localization-delocalization transition, one needs to
investigate the behavior of the 2D system in the thermodynamic limit. This, however, is not
possible in numerical methods, and therefore, we resort to a finite size scaling approach
in which we calculate the transmission while gradually increasing the size of the system for
a given amount of disorder. The result is then extrapolated to study the bulk behavior in
the thermodynamic limit.

Though there are many combinations possible for connecting the input and the output
leads with the cluster, for the work that follows, the leads are connected diagonally
with the clusters, since this arrangement generally gives higher transmission. In
addition to that, to minimize the effect of the boundary on the interior property of
the disordered clusters, we made good contacts by keeping the nine sites nearest to
both the input and the output contact points always occupied (that is, available to the
hopping particle). We calculate the transmission as a function of the system size
for many levels of disorder and at different energies. The general trend for all the
transmission curves at different energies are similar, so we will discuss here only
two energies, one that is away from the band center and one very close to the band center.

\subsection{Energies away from the band center}

We first study the transmission at energy $E = 1.6$ for different levels of disorder.
For each level of disorder, we have calculated the transmission by gradually increasing
the size of the clusters from $10 \times 10$ to a maximum size of $180 \times 180$.
We have randomly generated one thousand clusters of a given size for each level of
disorder and average transmission is calculated for each size of the cluster and
each level of disorder. The log-log plot of transmission against the size of the
clusters at $E = 1.6$ is shown for various disorder levels in the Fig.~\ref{tvsL1}.

\begin{widetext}
\begin{center}
\begin{figure}[h]
{\resizebox{6.5in}{5in}{\includegraphics{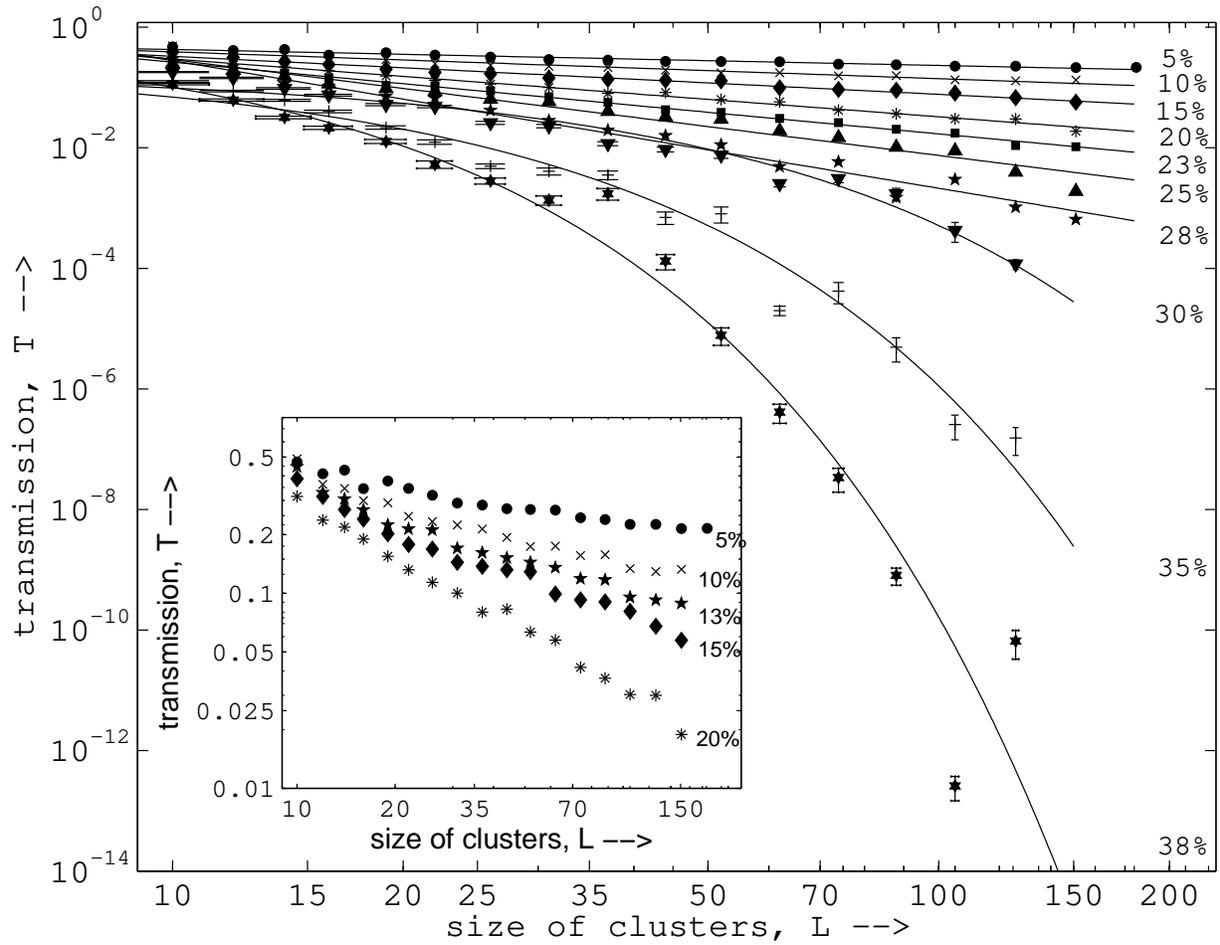}}}\\
\caption{The log-log plot of transmission through disordered clusters at $E = 1.6$.
         Each data point is the average over 1000 independent realizations. The
         transmission curves for $5\%$, $10\%$, $13\%$, $15\%$ and $20\%$
         disorders are separately shown in the inset.}
\label{tvsL1}
\end{figure}
\end{center}
\end{widetext}

We observe from Fig.~\ref{tvsL1} that at lower disorder transmission decreases almost linearly
in the log-log plot as the size of the clusters increases, suggesting a power-law
behavior of transmission in linear scale. This trend continues until disorder, q,
increases to about 28\%. Above that, transmission falls exponentially as is evident from
the lowest three curves in Fig.~\ref{tvsL1}. We have fitted the data both to power-laws
and exponentials and the best fit forms along with the corresponding fitting parameters
and coefficient of regression, $|R|^{2}$, for different levels of disorder are presented
in Table~\ref{tab1}. Although there are many data points that appear to be several
$\sigma$'s from the best fit exponentials particularly for higher disorder, they are
believed to be due to the discrete and loose-packed structure of the lattice and not
due to some unknown systematic errors.
(The transmission curves of the 13\% and 26\% disorder are not shown in the
Fig.~\ref{tvsL1} since they are very close to the nearby transmission curves).

\begin{table}[h]
\caption{Table for fitting parameters at $E = 1.6$. Shown in the parentheses are
         the lower and upper bounds for 95\% confidence level.}
\label{tab1}
\begin{tabular}{||c|c|c|c|c||}   \hline
   q    & Fit       & \multicolumn{2}{c|}{Parameters} & $|R|^{2}$ \\  \cline{3-4}
        & equation  & a                 & b                  &           \\  \hline
$5\%$   &           & 0.78 (0.69, 0.89) & 0.26 (0.23, 0.30)  & 0.95      \\  \cline{1-1} \cline{3-5}
$10\%$  &           & 1.09 (0.89, 1.32) & 0.44 (0.39, 0.50)  & 0.96      \\  \cline{1-1} \cline{3-5}
$13\%$  &           & 1.24 (1.01, 1.52) & 0.54 (0.49, 0.60)  & 0.97      \\  \cline{1-1} \cline{3-5}
$15\%$  & $T = a\cdot L^{-b}$ & 1.404 (1.16, 1.70) & 0.63 (0.58, 0.68)  & 0.98      \\  \cline{1-1} \cline{3-5}
$20\%$  &           & 2.70 (2.25, 3.25) & 0.96 (0.91, 1.01)  & 0.99      \\  \cline{1-1} \cline{3-5}
$23\%$  &           & 4.08 (3.47, 4.81) & 1.19 (1.15, 1.23)  & 0.99      \\  \cline{1-1} \cline{3-5}
$25\%$  &           & 11.3 (6.20, 20.8) & 1.59 (1.43, 1.75)  & 0.97      \\  \cline{1-1} \cline{3-5}
$26\%$  &           & 16.3 (8.98, 29.5) & 1.77 (1.61, 1.93)  & 0.97      \\  \cline{1-1} \cline{3-5}
$28\%$  &           & 36.2 (18.4, 71.5) & 2.12 (1.93, 2.30)  & 0.98      \\  \hline
$30\%$  &           & 0.18 (0.12, 0.25) & 0.06 (0.05, 0.06)  & 0.97      \\  \cline{1-1} \cline{3-5}
$35\%$  & $T = a\cdot \mbox{e}^{-bL}$ & 0.24 (0.12, 0.46)  & 0.12 (0.11, 0.13)  & 0.97      \\  \cline{1-1} \cline{3-5}
$38\%$  &           & 0.96 (0.15, 6.12) & 0.22 (0.19, 0.26)  & 0.94      \\  \hline
\end{tabular}
\end{table}

It should be noted that the curve fitting is performed in log-log scale and then
converted to linear scale along with the parameters. Thus to fit the data, say
for $q = 15\%$, we first calculate the logarithm of the size of the cluster and
the corresponding transmission, ie $x = log(L)$ and $y = log(T)$. We then fit
these logarithmic data with the curve, $y = m.x + c$, and converted the result
back to linear scale.

\begin{figure}[h]
{\resizebox{3.2in}{3.2in}{\includegraphics{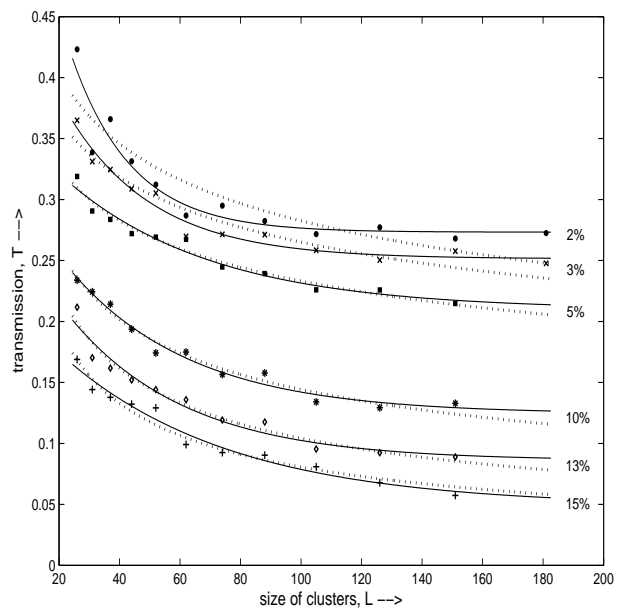}}}
\caption{The linear plot of transmission through clusters with 2\%, 3\%, 5\%, 10\%,
         13\% and $15\%$ disorders at $E = 1.6$ for larger sizes of the clusters
         ($L \geq 25$). Each data point is the average over 1000 independent realizations. The dotted lines
         represent power-law fits and the solid lines represent the exponential fits
         with offset.}
\label{tvsL2}
\end{figure}

Because of the large differences in transmission for different disorder, we used the
logarithmic scale for transmission in Fig.~\ref{tvsL1}, which in turn makes it difficult
to see the finer details of the transmission curve particularly at lower disorders.
For instance, Fig.~\ref{tvsL1} appears to suggest an excellent linear fit
at disorders below 15\%. To investigate more closely the nature of transmission
at lower disorder we have plotted the first five curves of Table~\ref{tab1} separately
in the inset of Fig.~\ref{tvsL1}. It is evident from the inset that transmission decreases
much more slowly than a power-law at larger values of $L$ for disorder up to about 13\%.
Above 13\% power-laws do appear to give good fits of the data for the entire range of
cluster sizes used. To investigate this lower disorder regime further, we have obtained
the transmission for 2\% and 3\% disorder and have plotted these data along with
those for 5\% to 15\% disorder in a linear graph as shown in Fig.~\ref{tvsL2} for
larger clusters ($L \geq 25$). Each data set is fitted with two different curves namely,

a power-law:
\begin{equation}
T = a\cdot L^{-b}
\label{eq6}
\end{equation}

and an exponential with offset:
\begin{equation}
T = a\cdot \mbox{e}^{-bL} + c
\label{eq7}
\end{equation}

The dotted and solid lines represent fits with Eq.~(\ref{eq6}) and (\ref{eq7}),
respectively. As we can see from Fig.~\ref{tvsL2}, exponential with offset, c,
gives significantly better visual fits for the data compared to power-laws
at least for 2\% and 3\% disorder, while the goodness of fits for higher disorders
appear to be a toss-up.

\begin{table}[h]
\caption{Table for fitting parameters of the linear plot at $E = 1.6$. Shown in the
         parentheses are the lower and upper bounds for 95\% confidence level.}
\label{tab2}
\begin{tabular}{||c|c|c|c|c|c|c|c|c|c||}   \hline
   q     & \multicolumn{4}{c|}{$T = a\cdot L^{-b}$}  & \multicolumn{5}{c||}{$T = a\cdot\mbox{e}^{-b L} + c$ } \\ \cline{2-10}
         & a     & b     & $|R|^{2}$ & SSE     & a    & b    & c    & $|R|^{2}$  & SSE       \\  \hline
   2\%   & 0.78  & 0.22  & 0.79      &$5\times$& 0.48 & 0.05 & 0.27       & 0.89 &$2\times$  \\
         &       &       &           &$10^{-3}$&      &      &(0.26, 0.29)&      &$10^{-3}$  \\  \hline
   3\%   & 0.67  & 0.20  & 0.91      &$1\times$& 0.27 & 0.03 & 0.25       & 0.96 &$5\times$  \\
         &       &       &           &$10^{-3}$&      &      &(0.24, 0.26)&      &$10^{-4}$  \\  \hline
   5\%   & 0.61  & 0.21  & 0.96      &$3\times$& 0.17 & 0.02 & 0.21       & 0.95 &$4\times$  \\
         &       &       &           &$10^{-4}$&      &      &(0.19, 0.23)&      &$10^{-4}$  \\  \hline
   10\%  & 0.78  & 0.37  & 0.97      &$3\times$& 0.21 & 0.02 & 0.12       & 0.97 &$3\times$  \\
         &       &       &           &$10^{-4}$&      &      &(0.11, 0.14)&      &$10^{-4}$  \\  \hline
   13\%  & 0.95  & 0.48  & 0.96      &$5\times$& 0.22 & 0.03 & 0.09       & 0.94 &$6\times$  \\
         &       &       &           &$10^{-4}$&      &      &(0.06, 0.1) &      &$10^{-4}$  \\  \hline
   15\%  & 1.00  & 0.55  & 0.96      &$4\times$& 0.18 & 0.02 & 0.05       & 0.96 &$4\times$  \\
         &       &       &           &$10^{-4}$&      &      &(0.02, 0.08)&      &$10^{-4}$  \\  \hline
\end{tabular}
\end{table}

To compare the goodness of fits we need to consider not only the visual fits but
the values of $|R|^{2}$ and SSE (sum square error) as well, and it is evident from
Table~\ref{tab2} and  Fig.~\ref{tvsL2}b that an exponential with offset gives
significantly better fits than a power law, except for 15\% disorder, at which
both fits are quite close.

The above study, thus, suggests the existence of three regimes of transmission. At
higher disorder transmission drops exponentially as the size of clusters increases.
At intermediate disorder, transmission follows power law, whereas at low disorder
transmission approaches a constant offset, c, suggesting possible delocalization of
the states. The value of the offset, however, depends on energy. For a given disorder,
c decreases as the energy increases, except for energies very close to the band center
where transmission itself is very low.

\subsection{Energies near the band center}

In this subsection we discuss the nature of transmission at an energy near the
band center. We investigate the scaling behavior at $E = 0.001$ for different
levels of disorder. The log-log plot of transmission as a function of system size is
shown in the Fig.~\ref{tvsL3}.

\begin{widetext}
\begin{center}
\begin{figure}[h]
{\resizebox{6.5in}{5in}{\includegraphics{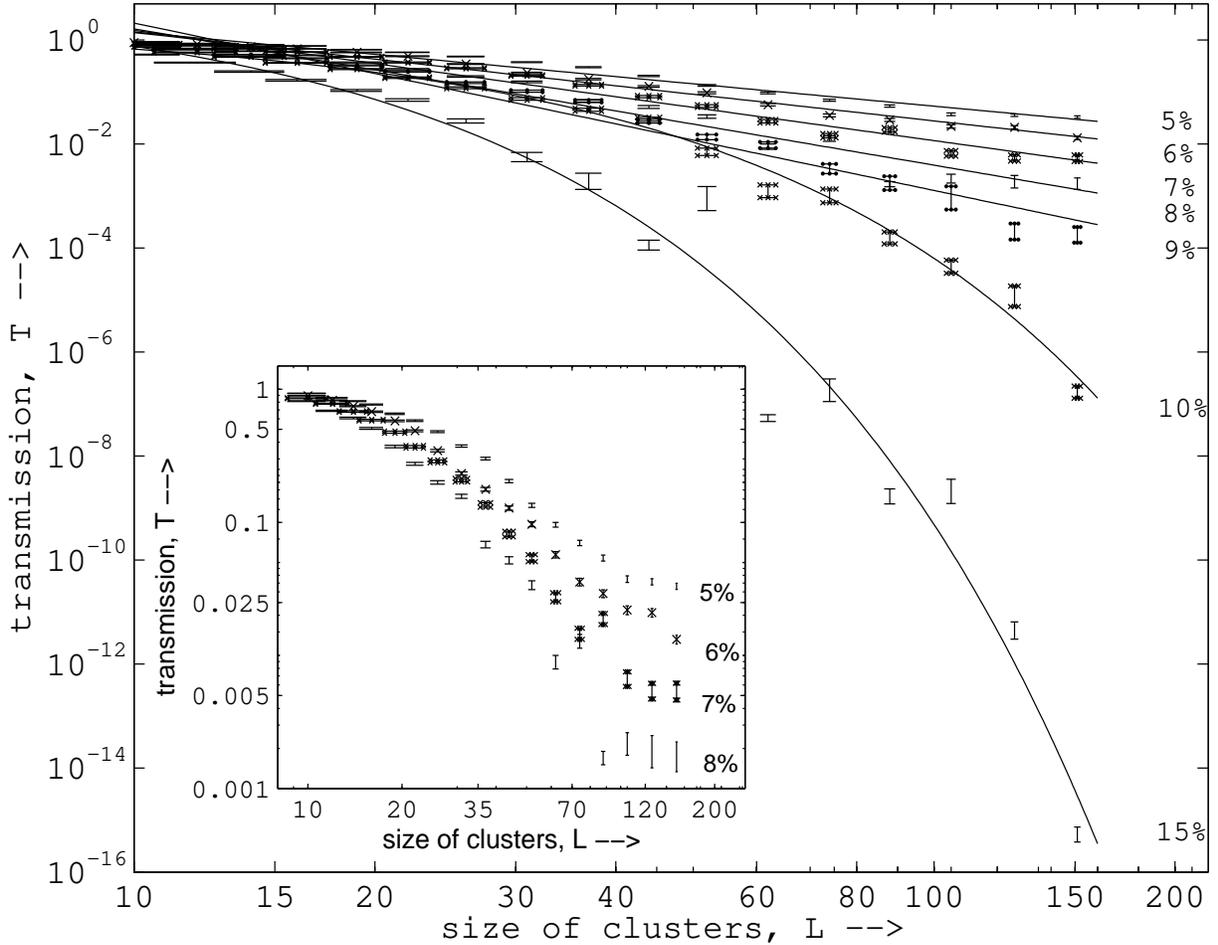}}}\\
\caption{The log-log plot of transmission through disordered clusters at $E = 0.001$.
         Each data point is the average over 1000 independent realizations. The
         transmission curves for $5\%$, $6\%$, $7\%$, and $8\%$ disorders are also
         separately shown in the inset.}
\label{tvsL3}
\end{figure}
\end{center}
\end{widetext}

At this energy, transmission decreases exponentially for disorder as low as 9\% as
is evident from Fig.~\ref{tvsL3}. Below 9\% the transmission curves appear to fit well with
power-laws. The fitting parameters of the different curves are tabulated in the
Table~\ref{tab3} along with the coefficient of regression, $|R|^{2}$.

\begin{table}[h]
\caption{Table for fitting parameters at $E = 0.001$. Shown in the parentheses are
         the lower and upper bounds for 95\% confidence level.}
\label{tab3}
\begin{tabular}{||c|c|c|c|c||}   \hline
    q   & Fit        & \multicolumn{2}{c|}{Parameters}        & $|R|^{2}$ \\  \cline{3-4}
        & equation   & a                 & b                  &           \\  \hline
$5\%$   &            & 38.2 (22.7, 64.5) & 1.43 (1.29, 1.57)  & 0.97      \\  \cline{1-1} \cline{3-5}
$6\%$   &            & 70.0 (69.9, 113)  & 1.70 (1.57, 1.83)  & 0.98      \\  \cline{1-1} \cline{3-5}
$7\%$   & $T = a\cdot L^{-b}$ & 204 (97.6, 429) & 2.12 (1.93, 2.32)  & 0.97  \\  \cline{1-1} \cline{3-5}
$8\%$   &            & 691 (216, 2206)   & 2.62 (2.31, 2.93)  & 0.96      \\ \hline
$9\%$   &            & 0.85 (0.53, 1.35) & 0.06 (0.06, 0.07)  & 0.96      \\ \cline{1-1} \cline{3-5}
$10\%$  &$T = a\cdot \mbox{e}^{-bL}$ & 1.88 (1.31, 2.69) & 0.10 (0.11, 0.10) & 0.99   \\  \cline{1-1} \cline{3-5}
$15\%$  &            & 8.00 (1.90, 33.7) & 0.24 (0.21, 0.26)  & 0.97      \\  \hline
\end{tabular}
\end{table}

The $|R|^{2}$ values for $5\%$, $6\%$, $7\%$ and $8\%$ disorder from the Table~\ref{tab3}
suggest a good fit of these data with power-laws. However, if we take a closer look at
the transmission curves for these disorders in the inset, we observe a significant
deviation from straight lines as the size of the cluster increases. It is clear that at
lower disorders, the falling trend in transmission is much slower than the power-law
similarly to what we have observed at energies far from the band center. However,
transmission is generally much smaller for the same amount of disorder compared with
the case of energies away from the band center.

\begin{figure}[h]
$\begin{array}{c}
{\resizebox{3.2in}{3.2in}{\includegraphics{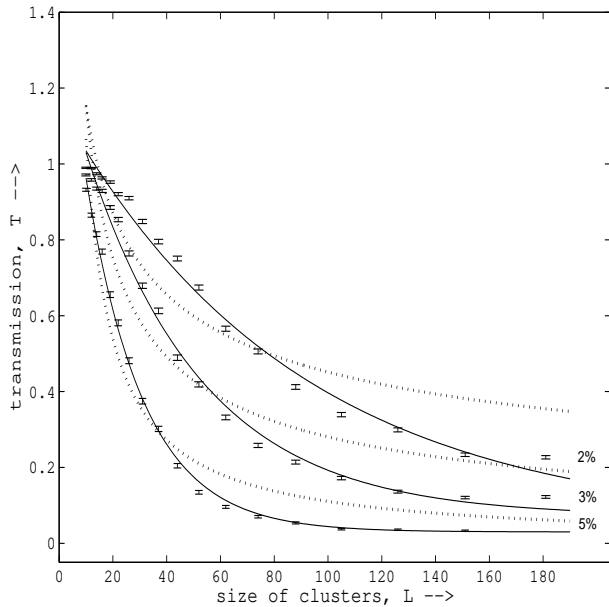}}}\\
(a)\\
{\resizebox{3.2in}{3.2in}{\includegraphics{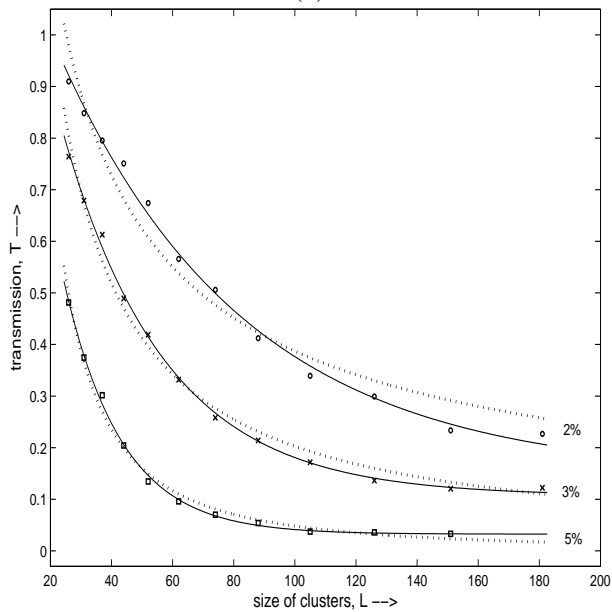}}}\\
(b)\\
\end{array}$
\caption{The linear plot of transmission through clusters with 2\%, 3\% and 5\%
         disorder at $E = 0.001$. (a) plot with all data points and (b) plot with
         data points only for larger sizes of the clusters. Each data point is the
         average over 1000 independent realizations. The dotted lines represent
         the power-law fits.}
\label{tvsL4}
\end{figure}

As before we have obtained two more transmission data for 2\% and 3\% disorder and the results
are plotted in Fig.~\ref{tvsL4}a along with that for 5\% disorder (the transmission
curve for 6\% disorder is too close to that of 5\% disorder and is not shown in the
the figure). We show in Fig.~\ref{tvsL4}a transmission data for the entire range of
cluster sizes. We have fitted these data with both a power-law and an exponential with
offset as before, and the best fit curves along with the fit parameters are
tabulated in the Table~\ref{tab4}. The dotted lines represent the power-law fits, and
the solid lines represent exponential fits with offset. We notice that at this energy the
best fit curve for each of the disorder amounts shown is clearly an exponential with offset
rather than a power-law. For more direct comparison with Fig.~\ref{tvsL2} where we considered
only the larger clusters ($L \geq 25$), we also show the corresponding figure and fits
for the same range of cluster sizes in Fig.~\ref{tvsL4}b as well. In the latter figure,
the difference in goodness of fits between power-law fits and exponential fits with offset
is not as large, but it is still clear that the latter fits the data better.

\begin{table}[h]
\caption{Table for fitting parameters of the linear plot at $E = 0.001$. Shown in
         the parentheses are the lower and upper bounds for 95\% confidence level.}
\label{tab4}
\begin{tabular}{||c|c|c|c|c|c|c|c|c|c||}   \hline
    q    & \multicolumn{4}{c|}{$T = a\cdot L^{-b}$}  & \multicolumn{5}{c||}{$T = a\cdot\mbox{e}^{-b L} + c$ } \\ \cline{2-10}
         & a     & b     & $|R|^{2}$ & SSE     & a    & b    & c    & $|R|^{2}$  & SSE       \\  \hline
   2\%   & 9.25  & 0.69  & 0.96      &$2\times$& 1.19 & 0.02 & 0.15       & 0.99 &$3\times$  \\
         &       &       &           &$10^{-2}$&      &      &(0.09, 0.20)&      &$10^{-3}$  \\  \hline
   3\%   & 22.7  & 1.03  & 0.99      &$7\times$& 1.44 & 0.03 & 0.11       & 0.99 &$1\times$  \\
         &       &       &           &$10^{-3}$&      &      &(0.09, 0.12)&      &$10^{-3}$  \\  \hline
   5\%   & 141   & 1.73  & 0.99      &$2\times$& 1.79 & 0.05 & 0.03       & 0.99 &$5\times$  \\
         &       &       &           &$10^{-3}$&      &      &(0.02, 0.04)&      &$10^{-4}$  \\  \hline
   6\%   & 198   & 1.95  & 0.99      &$4\times$& 1.53 & 0.06 & 0.02       & 0.99 &$7\times$  \\
         &       &       &           &$10^{-4}$&      &      &(0.01, 0.03)&      &$10^{-4}$  \\  \hline
\end{tabular}
\end{table}

The issue of localization length, $L_l$, must also be discussed. Our finite size scaling
approach does not rely on any single system size but rather focuses on the trend as
it increases. In fact this analysis detects different trends depending on the dilution
amount and we can estimate the localization length from the data themselves at least
in the clearly observed exponentially localized regime. Table~\ref{tab5} summarizes
the simulation estimates of $L_l$ at different energies in the regions of disorder where
the respective exponential fits ($a\cdot\mbox{e}^{-bL}$) are significantly better than
other types of fits. The localization length in the exponential regime can be estimated
from $L_l \sim b^{-1}$. It is evident that most of our system sizes are sufficiently large
compared with these estimates of $L_l$ and thus our results are internally consistent
with exponential localization at these amounts of disorder.

\begin{table}[h]
\caption{Estimated values of the localization lengths from $T = a\cdot \mbox{e}^{-bL}$,
         localization length being $b^{-1}$}
\label{tab5}
\begin{center}
\begin{tabular}{||c|c|c|c|c|c|c|c||}   \hline
q    & \multicolumn{7}{c||}{Localization length (in units of lattice constant} \\ \cline{2-8}
     & $E=$   & $E=$    & $E=$    & $E=$    & $E=$   & $E=$   & $E=$    \\
     & 0.001  & 0.05    & 0.5     & 1.0     & 1.2    & 1.6    & 1.9     \\ \hline
 9\% & 15.63  & -       & -       & -       & -      & -      & -       \\ \hline
10\% & 9.71   & -       & -       & -       & -      & -      & -       \\ \hline
15\% & 4.26   & -       & -       & -       & -      & -      & -       \\ \hline
20\% & 2.69   & 16.67   & -       & -       & -      & -      & -       \\ \hline
25\% & -      & 6.29    & -       & -       & -      & -      & 27.03   \\ \hline
30\% & -      & -       & -       & 19.30   & -      & 17.24  & 13.26   \\ \hline
31\% & -      & -       & 13.89   & -       & 16.69  & -      & -       \\ \hline
32\% & -      & -       & 13.89   & -       & -      & -      & -       \\ \hline
35\% & -      & -       & 8.85    & 6.75    & 7.12   & 8.16   & 5.37    \\ \hline
38\% & -      & -       & 4.85    & 4.32    & 4.14   & 4.44   & -       \\ \hline
\end{tabular}
\end{center}
\end{table}

\begin{figure}[h]
{\resizebox{3.2in}{3.2in}{\includegraphics{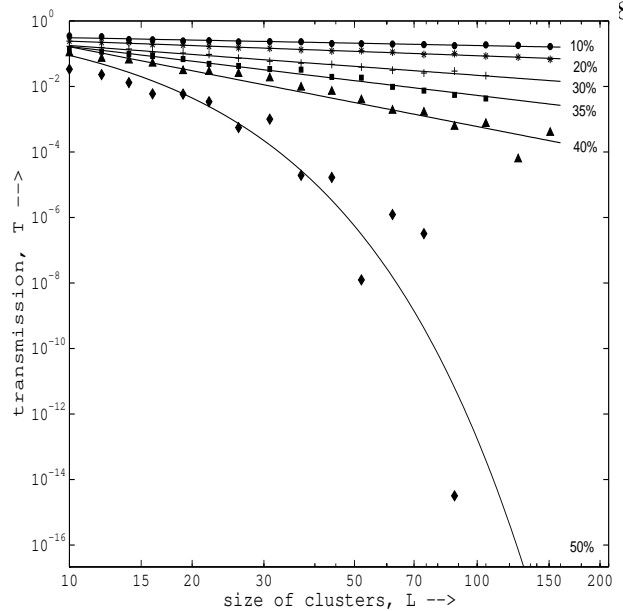}}}\\
\caption{The log-log plot of transmission through disordered clusters in triangular
         lattice at $E = 1$. Each data point is the average over 1000 independent
         realizations.}
\label{tvsL9}
\end{figure}

For completeness we present here the scaling result for triangular lattice only for
energy $E = 1$ for different disorder. The curves appear very similar to those for square
lattice as is evident from the Fig.~\ref{tvsL9} except that the exponentially localized
regime appears at a higher disorder. The result is expected since triangular lattice
has six nearest neighbors and therefore requires more disorder to reduce the
transmission.

\section{Summary and Conclusion}

In this paper we have studied the behavior of a quantum particle in two dimensional
disordered clusters in quantum percolation model. Our approach is based on calculating
the transmission of the particle that enters into the cluster through a one dimensional
lead at one side of the cluster and exit through another lead at the other side of
the cluster.

Our study based on finite size scaling suggests the existence of three different
regimes depending on disorder. The range of these regimes, however, depends on the
energy of the particle. Although we showed in previous sections results from two
representative energies ($E = 1.6$ that is far from the band center and $E = 0.001$
that is very close to it), we have in fact obtained and analyzed the data for many
more values of the energy. Table~\ref{tab6} shows our approximate estimates of the range
of disorder for which these regimes exist as the energy is varied.

\begin{table}[htbp]
\caption{Classification of different regime in 2D disordered system.}
\label{tab6}
\begin{tabular}{||c|c|c|c||}   \hline
Energy     & \multicolumn{3}{c||}{Range of disorder} \\ \cline{2-4}
           & delocalized        & power law         & exponential    \\
           & states             & localization      & localization   \\
           & regime             & regime            & regime         \\  \hline
0.001      & $0 - 6\%$          & $7\% - 8\%$       & $\geq 9\%$     \\  \hline
0.05       & $0 - <15\%$        & $>15\% - <20\%$   & $\geq 20\%$    \\  \hline
0.5        & $0 - 15\%$         & $>15\% - 31\%$    & $\geq 32\%$    \\  \hline
1.0        & $0 - 15\%$         & $>15\% - 28\%$    & $\geq 30\%$    \\  \hline
1.2        & $0 - 15\%$         & $>15\% - 30\%$    & $\geq 31\%$    \\  \hline
1.6        & $0 - <15\%$        &$\geq 15\% - 28\%$ & $\geq 30\%$    \\  \hline
1.9        & $0 - <15\%$        & $>15\% - 23\%$    & $\geq 25\%$    \\  \hline
\end{tabular}
\end{table}

Our study, thus, suggests that at lower disorder states are delocalized, contrary to
the one parameter scaling theory of Abrahams {\it et al} \cite{abrahams:79}. In
addition to delocalized transition we observe a second kind of phase transition
between the power-law to exponentially localized regimes.

For all energies considered in this work, except for those close to $E = 0$, the
delocalized states appear if the disorder is less than about 15\%. Above 15\%, there
exist two different localization regimes. For intermediate energies ($0.5 < E < 1.6$),
states of the particle show a weaker form of localization characterized by a power-law
size dependence of transmission for the
range of disorder between 15\% to $\sim 30\%$. Above this range all states are
localized exponentially. For energies close to $E = 2$, the width of this power-law
localization regime is reduced and states become exponentially localized even
for 25\% disorder.

The transmission near the band center, however, differs in some important ways compared
to that at other energies. At $E \sim 0$, transmission is very low in the thermodynamic
limit even for very low disorder. Though it is difficult to locate the delocalization
transition because of the low transmission, our finite size analysis does show a behavior
much weaker than power-law localization (inset of Fig.~\ref{tvsL3}). The width of the
power-law localization regime is very small and exponentially localized states appear for
disorder as low as 9\%.

\section*{Acknowledgements}

We would like to thank A. Overhauser, Y. Lyanda-Geller, S. Savikhin,and many others for
discussions and the Physics Computer Network (PCN) at Purdue University where most of
the numerical work was performed.

\end{document}